\documentclass[longbibliography,superscriptaddress,showpacs,aps,twocolumn,prb,floatfix]{revtex4-2}
\usepackage{amssymb}
\usepackage{amsmath}
\usepackage{graphicx}
\usepackage{bm}

\usepackage[caption=false]{subfig}
\usepackage[colorlinks=true,linkcolor=blue,citecolor=blue,urlcolor=blue]{hyperref}

\begin{document}

\title{Role of asymmetry in thermoelectric properties of a double quantum dot out of equilibrium}

\author{D. Perez Daroca}
\affiliation{Gerencia de Investigaci\'on y Aplicaciones, GAIDI, CNEA, 1650 San Mart\'{\i}n, Buenos Aires, Argentina}
\affiliation{Consejo Nacional de Investigaciones Científicas y Técnicas, 1025 CABA, Argentina}

\author{P. Roura-Bas}
\affiliation{Centro At\'{o}mico Bariloche, GAIDI,  8400 Bariloche, Argentina}
\affiliation{Consejo Nacional de Investigaciones Científicas y Técnicas, 1025 CABA, Argentina}

\author{A. A. Aligia}
\affiliation{Instituto de Nanociencia y Nanotecnolog\'{\i}a 
CNEA-CONICET, GAIDI,
Centro At\'{o}mico Bariloche, 8400 Bariloche, Argentina}

\begin{abstract}
We investigate the thermoelectric properties of a double quantum dot system coupled to two metallic reservoirs, focusing on two main effects: (i) the influence of coupling asymmetry between the quantum dot and the reservoirs on the Seebeck coefficient, and (ii) the impact of asymmetry in the energy levels of the dots on current rectification. In the first case, we find that introducing moderate asymmetry significantly enhances the Seebeck coefficient. In the second case, while rectification vanishes when the energy levels are degenerate, substantial rectification is achieved when one energy level lies below and the other above the Fermi level. We further interpret the dependence of rectification magnitude and shape on system parameters using analytical results from a spinless model.
\end{abstract}

%\pacs{73.23.-b, 71.10.Hf, 75.20.Hr}

\maketitle

\section{Introduction}

\label{intro}

During this century, the study of thermoelectric transport in nanoscale systems has advanced notably due to possible applications and the need for efficient energy conversion technologies. Relevant experiments can be found in Refs. \cite{guo13,kim15,rincon,cui17,miao18,cui19,erdman,svilans,josef,dutta}. The corresponding calculations use, in general, a multilevel single impurity Anderson model to describe molecules or quantum dots (QDs) \cite{bene,cui17,erdman,boese,hump02,kra07,Kubala,pola,costi,leij,azem,see,ng,azem2,dorda16,Ludovico16,Romero17,sierra17,li17,Dare17,li18,burkle,asym,KK,tesser,mana,mina,cortes,mana24,Deghi24}. QDs have the advantage of their highly tunable electronic properties.

Recently, double quantum dots (DQDs) have received much attention because they offer additional control knobs like 
two on-site energies, interdot hopping and interdot Coulomb repulsion  \cite{tesser,donsa,Dare17,craven,sierra,dare,yada,heat,lava20,lomba,zhang,ghosh,mana2,daroca23,sobrino24}. Additionally, DQDs have been considered for a variety of effects and applications, including qubits, quantum interference, thermoelectric devices, and more. From an experimental perspective, DQDs have been realized in systems such as AlGaAs/GaAs heterostructures \cite{keller}, Si-MOS devices \cite{king}, InAs/InP nanowires\cite{dorsch} and Ge hut nanowires \cite{zhou}, to name a few.

The more robust technique used for one QD is the numerical 
renormalization group (NRG) used mainly at equilibrium 
\cite{dare,donsa,costi,anders08,Ngh18,mana,mina,mana2,mana22,mana24}, but interesting developments were made for the non-equilibrium case \cite{anders08,Ngh18,mana22,mana24}.
Other techniques employed  include the equations of motion approach 
\cite{pola,li18,rome09,rapha,rome,crep,sobrino24}, renormalized perturbation theory in the on-site Coulomb repulsion $U$ or
similar Fermi-liquid methods \cite{ng,asym,KK,heat,oguri01,oguhe,mora09,sela,ct,karki18,tera20,tera}, slave-boson
mean-field approximation \cite{sierra17,sierra,geme,dong02,hamad13,aguado,rosa02}, non-crossing approximation 
(NCA) \cite{lomba,win,hettler,sitri1,sitri2,serge,tetta,vibra,desint19,choi,roura10} or functional renormalization-group
restricted to large bias voltages or magnetic fields \cite{rosch03}. The limitations of these approaches were explained in our previous work \cite{daroca23}.

In this work we use the NCA. For the simplest Anderson model, 
this approach is shown to accurately reproduce NRG results for the spectral density at temperatures $T>0.1T_K$, where $T_K$ is the Kondo temperature \cite{costi96}. It was also  successfully used to calculate the non-equilibrium properties of different systems \cite{hettler}, including two-level quantum dots and  C$_{60}$ molecules displaying a quantum phase transition \cite{sitri1,sitri2,serge}, a nanoscale Si transistor \cite{tetta}, and vibrating molecules \cite{vibra,desint19}, among others \cite{lomba,choi}. It also reproduces correctly the scaling of the conductance for small bias voltage $V$ and temperature $T$ \cite{roura10}.
More recently, the NCA was used to study 
a three-terminal thermoelectric engine, for energy-harvesting purposes \cite{Dare17} and to control the thermopower among other properties \cite{lomba}.

Previously, we employed the NCA method to calculate the thermoelectric properties of a symmetric DQD, demonstrating its advantages over alternative approaches. Building on earlier findings, which indicate that thermoelectric power increases with asymmetry in the coupling to the leads \cite{asym,mana}, and that energy asymmetry in the dots can result in significant rectification effects \cite{heat}, this work explores the impact of these asymmetries on the  DQD system.

The structure of this paper is as follows. In Sec. \ref{model}, we present the model, describe the methods, and derive the expressions for the electrical and heat currents. Thermoelectric power results are discussed in Sec. \ref{seeb}, followed by the analysis of rectification in Sec. \ref{rect}. Finally, Sec. \ref{sum} provides a summary and discussion of the key findings.

\section{Model and methods}

\label{model}

\begin{figure}[h!]
\centering
\includegraphics[width=0.45\textwidth]{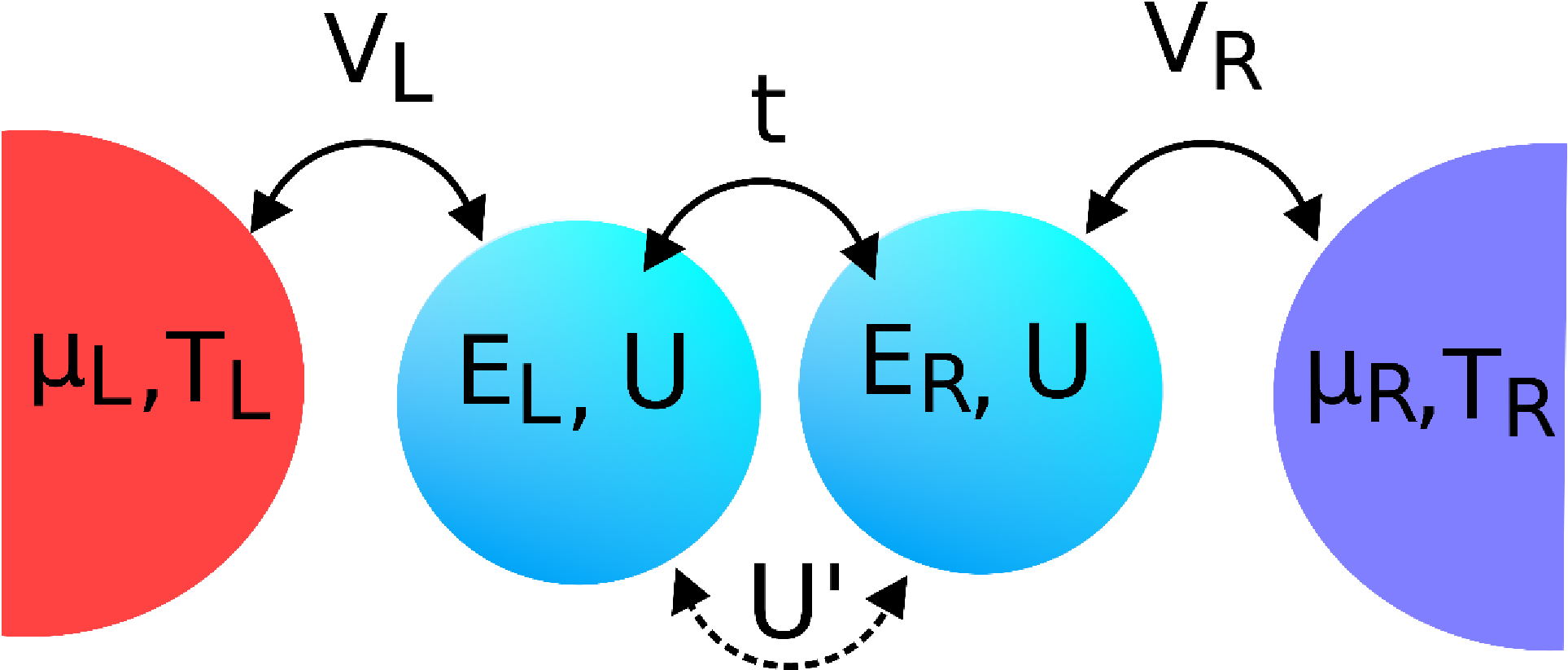}  
\caption{(Color online) Schematic representation of the system analyzed in this work: two quantum dots (circles) connected in series to two conducting leads (semicircles) with differing temperatures and chemical potentials.}
\label{scheme}
\end{figure}

Figure \ref{scheme} displays the diagram of the system in a nonequilibrium state. The setup includes a double quantum dot, where each dot is tunnel-coupled to a metallic reservoir. The system is described by the following Hamiltonian:

\begin{equation}
H=H_{DQD}+H_{c}+H_{V}.  \label{ham}
\end{equation}

The first term describes the DQD, 
\begin{eqnarray}
H_{DQD} &=&\sum_{\nu \sigma }E_{\nu }d_{\nu \sigma }^{\dagger }d_{\nu \sigma
}+\sum_{\nu }Un_{\nu \uparrow }n_{\nu \downarrow }  \notag \\
&&+U^{\prime }\sum_{\sigma \sigma ^{\prime }}n_{L\sigma }n_{R\sigma ^{\prime
}}\\
	&&-t\sum_{\sigma }\left( d_{L\sigma }^{\dagger }d_{R\sigma }
+\text{H.c.}\right), \notag  \label{hdqd}
\end{eqnarray}
where $E_{\nu }$ and $U$ ($U^{\prime }$) are the energy levels and the intra-
(inter-) dot Coulomb repulsion, respectively,  $\nu =\{L,R\}$ labels the left
and right dots (and leads), and $\sigma ={\uparrow ,\downarrow}$ stands for
the spin projection. The hopping energy between dots is represented by $t$.

The second term describes the conducting leads 
\begin{equation}
H_{c}=\sum_{\nu}H_{\nu}=\sum_{k\nu \sigma }\varepsilon _{\nu k\sigma }\,c_{\nu k\sigma
}^{\dagger }c_{\nu k\sigma },  \label{hleads}
\end{equation}
and the last one, describes the hybridization between each dot and its
respective lead 
\begin{equation}
H_{V}=\sum_{k\nu \sigma }\left( V_{k\nu }\,d_{\nu \sigma }^{\dagger }c_{\nu
k\sigma }+\text{H.c.}\right) ,  \label{hyb}
\end{equation}
where $V_{k\nu}$ describes the hopping elements between the leads and the QDs.

To address the system within the NCA and avoid the need for vertex corrections, we diagonalize $H_{DQD}$ and retain only two neighboring configurations, which correspond to the $U \rightarrow +\infty$ limit. This simplifies the problem to a multilevel system connected to two conducting reservoirs. By exploiting electron-hole symmetry, two distinct scenarios arise in the DQD:  fluctuations between zero and one particles, and between one and two particles. Therefore, it is sufficient to focus on the eigenstates and energies of $H_{DQD}$ for one and two particles.

Throughout the results focused on the thermoelectric power (Sec. \ref{seeb}), we assume that the DQD has inversion symmetry, as, in this case, the effect of asymmetry is accounted for in the couplings with the leads. Therefore,  $E_{L}=E_{R}\equiv E_{d}$. The eigenstates for the one-particle sector are the even and odd linear combinations

\begin{eqnarray}
d_{e\sigma }^{\dagger } &=&\frac{1}{\sqrt{2}}\left( d_{L\sigma }^{\dagger
}+d_{R\sigma }^{\dagger }\right) ,\text{ }E_{e}=E_{d}-t,  \notag \\
d_{o\sigma }^{\dagger } &=&\frac{1}{\sqrt{2}}\left( d_{L\sigma }^{\dagger
}-d_{R\sigma }^{\dagger }\right) ,\text{ }E_{o}=E_{d}+t.  \label{deo}
\end{eqnarray}

It is straightforward to observe that, in the new basis, the problem of fluctuations between zero and one particles takes the same form as the previously studied interference case, with a level splitting of $\delta = 2t$ \cite{desint11,desint19,benz}.

In the two-particle sector, the four relevant eigenstates for $U-U^{\prime } \rightarrow +\infty$ consist of an even singlet and an odd triplet. Using the notation $|Sm\rangle$ for the total spin and its projection, these states are

\begin{eqnarray}
|00\rangle  &=&\frac{1}{\sqrt{2}}\left( d_{L\uparrow }^{\dagger
}d_{R\downarrow }^{\dagger }-d_{L\downarrow }^{\dagger }d_{R\uparrow
}^{\dagger }\right) |0\rangle, \notag\\
|11\rangle  &=&d_{L\uparrow }^{\dagger }d_{R\uparrow }^{\dagger }|0\rangle ,\\
|10\rangle  &=&\frac{1}{\sqrt{2}}\left( d_{L\uparrow }^{\dagger
}d_{R\downarrow }^{\dagger }+d_{L\downarrow }^{\dagger }d_{R\uparrow
}^{\dagger }\right) |0\rangle ,  \notag \\
|1-1\rangle  &=&d_{L\downarrow }^{\dagger }d_{R\downarrow }^{\dagger
}|0\rangle .  \notag
	\label{states}
\end{eqnarray}
The energy of the degenerate triplet is given by $E_{1} = 2E_{d} + U^{\prime}$, while the energy of the singlet is $E_{0} = E_{1} - J$, where $J = 4t^{2}/(U - U^{\prime})$. Note that, in the present limit, $J\rightarrow0$, $E_{0} = E_{1}$. For a detail of the effective Hamiltonian in the new basis and a brief explanation of the NCA treatment, see Ref. \cite{daroca23}.

Here, we outline the conventions and expressions for the charge and heat currents required to calculate the thermoelectric power and rectification. Treating the system depicted in Fig. \ref{scheme} as an interacting region connected to conducting leads, the charge and energy currents are expressed as $J^{\nu}_{C} = -e\langle \dot{N}_{\nu} \rangle$, $J^{\nu}_{E} = -\langle \dot{H}_{\nu} \rangle$, and the heat current as $J^{\nu}_{Q} = J^{\nu}_{E} - \mu_{\nu} J^{\nu}_{C}$, where $e$ is the elementary charge and ${N}_{\nu} = \sum_{k \sigma} c^{\dagger}_{\nu k \sigma} c_{\nu k \sigma}$. Current conservation leads to $J^{L}_{C,E} = -J^{R}_{C,E}$, and if $\mu_R = \mu_L$, then $J^{L}_{Q} = -J^{R}_{Q}$. In the case of the rectification results, we assume $\mu_R = \mu_L = 0$. Therefore, the term proportional to $\mu_{\nu}$ vanishes, resulting in $J^{\nu}_{Q} = J^{\nu}_{E}$. Moreover, once the appropriate sign for each flow direction is chosen, the index $\nu$ can be omitted from the definitions of the currents. We adopt the convention that currents flowing from the left lead to the right are positive. 

Using a procedure similar to the one detailed in the Appendix of Ref. \onlinecite{benz}, the charge and heat currents can be expressed in terms of the physical Keldysh Green's functions derived from the NCA approximation

\begin{eqnarray}\label{mw-charge-heat-currents}
 J_{C}&=&\frac{ie}{h}\int d\omega~\mbox{Tr}\Big[ \big( \mathbf{\Gamma^{L}} f_{L}(\omega)-\mathbf{\Gamma^{R}} f_{R}(\omega) \big)\mathbf{G^{>}_{d}}(\omega)\notag \\  &&+\big( \mathbf{\Gamma^{L}} f_{L}(-\omega)-\mathbf{\Gamma^{R}} f_{R}(-\omega) \big)\mathbf{G^{<}_{d}}(\omega)\Big],\\
 J_{Q}&=&\frac{ie}{h}\int d\omega~\omega\mbox{Tr}\Big[ \big( \mathbf{\Gamma^{L}} f_{L}(\omega)-\mathbf{\Gamma^{R}} f_{R}(\omega) \big)\mathbf{G^{>}_{d}}(\omega)\notag \\  &&+\big( \mathbf{\Gamma^{L}} f_{L}(-\omega)-\mathbf{\Gamma^{R}} f_{R}(-\omega) \big)\mathbf{G^{<}_{d}}(\omega)\Big],
	\label{jQ}
 \end{eqnarray}
where 
\begin{equation}
  \mathbf{\Gamma^{L}}=\Gamma_{L}\left(
    \begin{array}{cc}
      1&1\\
      1&1
    \end{array}
  \right),\quad
  \mathbf{\Gamma^{R}}=\Gamma_{R}\left(
    \begin{array}{cc}
      1&-1\\
      -1&1
    \end{array}
  \right),
\end{equation}
are the matrices that couple the even and odd levels to the left and right reservoirs. Additionally, $\Gamma_{\nu} = \pi V_{\nu}^2 / D$, where $2D$ is the width of the conduction band, and $f_{\nu}(\omega) = \left[ 1 + \exp\left(\frac{\omega - \mu_{\nu}}{k_B T_{\nu}}\right)\right]^{-1}$ represents the Fermi function. It is important to note that since $\mathbf{\Gamma^{L}}$ and $\mathbf{\Gamma^{R}}$ are not proportional, it is impossible to eliminate the non-equilibrium Green's functions from the current expressions. Even in the linear response regime, the full non-equilibrium formalism is required to compute the electric and thermal conductances by numerically deriving the corresponding currents.

In Sec. \ref{rect}, we analyze rectification in the one-particle sector by relaxing the assumption of inversion symmetry. Consequently, $E_R$ is no longer equal to $E_L$, leading to the following modifications in the eigenenergies described by Eqs. (\ref{deo}):

\begin{eqnarray}
	E_o=\overline{E}+\Delta E,\\
	E_e=\overline{E}-\Delta E,\\
	\overline{E}=(E_L+E_R)/2,\\
	\Delta E=\sqrt{\big (\frac{E_L-E_R}{2}\big)^2+t^2. }
\label{deoRL}
\end{eqnarray}

Similarly, the coupling matrices  transform in the following manner:

\begin{equation}
  \mathbf{\Gamma^{L}}=\Gamma_{L}\left(
    \begin{array}{cc}
      u^2&uv\\
      uv&v^2
    \end{array}
  \right),\quad
  \mathbf{\Gamma^{R}}=\Gamma_{R}\left(
    \begin{array}{cc}
      v^2&-uv\\
      -uv&u^2
    \end{array}
  \right),
\end{equation}
where 
\begin{equation}
	u=\frac{1}{\sqrt{1+\big(\frac{E_e-E_L}{t}\big)^2}}
\end{equation}

and
\begin{equation}
	v=\Big(\frac{E_e-E_L}{t}\Big)u .
\end{equation}

\section{Results}

\label{res}

\subsection{Thermoelectric power}
\label{seeb}

In this section, we present the results for the thermoelectric power, also known as the Seebeck coefficient. This coefficient is defined as:

\begin{equation}\label{seebeck}
 S = - \frac{d(\Delta V)}{d(\Delta T)}\big\vert_{J_C = J_0} = \frac{\partial J_C}{\partial(\Delta T)} \Big/
                                                                 \frac{\partial J_C}{\partial(\Delta V)},
\end{equation}
where the condition $J_C = J_0$ allows for the calculation to be performed at a non-zero voltage, leading to a finite charge current $J_0$, referring to the differential Seebeck coefficient in this case.

The effect of asymmetry in the coupling to the leads on the thermoelectric power is analyzed by defining an asymmetry factor, $\alpha = \Gamma_L/\Gamma_R$.

\subsubsection{Fluctuations between 0 and 1 particles in the system}

\label{see01}

In this subsection we assume $U^\prime$ to be sufficiently large to avoid occupancy of more than
one particle in the double-dot system. The case in which the 
total number of particles in the system fluctuates between three and four particles can be mapped to the specific case treated here by an electron-hole transformation \cite{daroca23}. This transformation changes the sign of the
thermopower but not its magnitude.

We set the Fermi energy $\epsilon_F = 0$ as the reference point for one-particle energies. This transforms the problem into one of transport through a single dot or molecule with two levels, where destructive interference influences transport \cite{desint19,desint11,benz}. In the limits of $t \rightarrow 0$ and $E_L \rightarrow E_R$, the spectral density of each dot converges to that of the SU(4) Anderson model, which incorporates both spin and ''orbital" degeneracy. This model is characterized by a narrow peak of half width 
of the order of the Kondo temperature $T_K^{\text{SU(4)}}$
centered above the Fermi level by an energy also on the scale of $T_K^{\text{SU(4)}}$. In addition, there is a broader charge-transfer peak located at $E_d$, with a width of approximately $4\Delta$ \cite{anchos}, where $\Delta$ represents the resonant level width \cite{restor}.

We define $\Gamma = \Gamma_L + \Gamma_R = 1$ as the unit of energy and take $E_d = -4$. The set of parameters used in this work is a subset of a larger set; however, within this subset, we find the most interesting results. We also take units such that the Boltzmann constant $k_B=1$.

The variational equation \cite{su42,desint11} used to calculate $T_K$ is given by

\begin{equation}
T_{K}=\left\{ (D+\delta )D\exp \left[ \pi E_{d}/(2\Delta )\right]
+\delta ^{2}/4\right\} ^{1/2}-\delta /2.  
\label{tkd}
\end{equation}

\begin{figure}[ht]
\centering
\includegraphics[width=1.0\columnwidth]{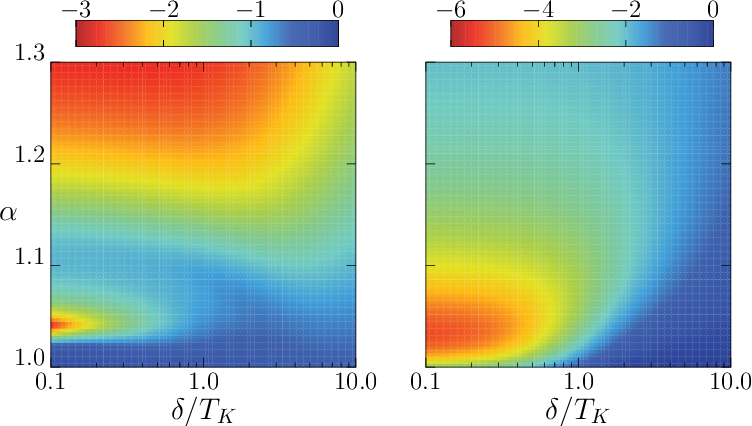}
\caption{(Color online) Seebeck coefficient as a function of 
$\alpha$ and $\delta/T_K$ for $T=0.1 T_K$ and two values of $V$. 
Left: $V=-2 T_K$. Right:  $V=2 T_K$.}
\label{seef}
\end{figure}

The ratio $\delta/T_K$ used in Fig. \ref{seef}, is calculated using Eq. (\ref{tkd}). For instance, $\delta = 0.001$ and $T_K = 0.01$ lead to $\delta/T_K = 0.1$.

We  looked for large values of the thermoelectric power $|S|$  for different asymmetry of the couplings $\alpha$, voltage $V$, temperature $T$, and splittings $\delta$. In Fig. \ref{seef} we show contour plots containing the largest values of  $|S|$ that we find. Specifically, for $T=0.1 T_K$, $V=-2 T_K$, $\delta$ smaller or of the order of $T_K$ and moderate $\alpha$, values of
$S \sim -3/e$ are obtained. Changing the voltage to $V=2 T_K$, results in even larger absolute values for small 
$\alpha$, but in a more restrictive zone of parameters. For $V=0$ (not shown) the maximum $|S| \sim 1.1/e$ is obtained for $\alpha=1$ and $\delta=0$. 

In general, these values are remarkably large. In previous
studies usually the largest values of $|S|$ are of the order of $1/e$ \cite{costi,lava20,lomba}.

We find that increasing $T$ to $T_K$, the values of $|S|$ are, in general, below $1/e$ except for $V=-2 T_K$, $\alpha \sim 1$ and
$\delta < 0.6 T_K$ for which 
$S \sim -1.8/e$. For $T=10 T_K$ $|S| \sim 0.4$ in the range of voltages studied ($-2 T_K \leq V\leq 2 T_K$).

The decrease in the absolute value of the thermopower as $\delta$ and $T$ increase is expected. For $V=0$, small $T$ and $\delta=0$, the spectral density tends to that corresponding to the SU(4) Anderson model in the 
Kondo limit, which is asymmetric and lies above the Fermi energy
\cite{su42,restor} favoring a negative and large thermopower \cite{see}.
As $\delta$ increases, there is a crossover to the SU(2) regime 
\cite{su42} with a more symmetric spectral weight and $|S|$ decreases.
Increasing the temperature, the Kondo peak is reduced and with it $|S|$.
For large temperature the thermopower is dominated by the charge-transfer
peak below the Fermi energy \cite{anchos} and $S$ becomes positive \cite{see}.

While for one dot, the effect of increasing $|S|$ with $\alpha$
was clear because it increases the asymmetry of the spectral density 
in the SU(2) regime \cite{asym,mana}, in our case its effect seems
to compete
with the natural asymmetry in the SU(4) regime for small $\delta$.

\subsubsection{Fluctuations between 1 and 2 particles in the system}

\label{see12}

The problem for fluctuations between two and three particles can be mapped into the corresponding one for one and two particles  using the electron-hole transformation mentioned at the beginning of Sec. \ref{see01}. In the following we refer to the latter case.

\begin{figure}[hb]
\centering
\includegraphics[width=1.0\columnwidth]{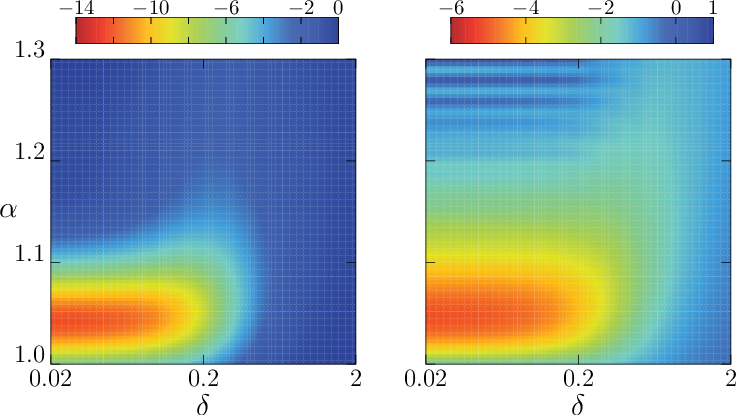}
\caption{(Color online) Seebeck coefficient as a function of 
$\alpha$ and $\delta$ for $T=0.1 T_K$ and two values of $V$. 
Left: $V=-2 T_K$. Right:  $V=2 T_K$.}
\label{seef2}
\end{figure}

We  choose $E_d=-4$, $U^\prime=0$, and $U-U^{\prime } \rightarrow +\infty$. We compute $T_K$ from the equilibrium conductance of an equivalent SU(4) Anderson model. Specifically, $T_K$ is defined as the temperature such that $G(T)=G_0/2$ with $G_0=2e^2/h$. Two interesting contour plots in the plane of asymmetry $\alpha$ and  $\delta$ are shown in Fig. \ref{seef2} for $T=0.1 T_K$. For both negative and positive $V$, a region can be identified between $0.02 < \delta < 0.2$ and $1 < \alpha < 1.1$, where $|S|$ reaches its maximum. However, in the negative $V$ case, this maximum is more than twice as large, reaching considerable high values ($ \sim 14/e$). 

In both scenarios, increasing either the asymmetry or the hopping leads to a substantial reduction in $|S|$. For moderate temperatures, $0.77T_K$, the negative voltage $V$ continues to exhibit the highest Seebeck coefficients in the same region, with maxima reaching $1.7/e$. At high temperatures, $10T_K$, the Seebeck coefficient $|S|$ decreases significantly and shows almost no dependence on $V$ (not shown). Due to the competition of several physical factors, we do not have a clear explanation for this behavior.

\subsection{Rectification}
\label{rect}

The rectification of thermal currents is defined as

\begin{equation}\label{r}
\mathcal{R}=\Big\vert \frac{J_{+}-J_{-}}{J_{+}+J_{-}}\Big\vert,
\end{equation}
 with $\mathcal{R}=1$ being the upper bound, where $J_{+}$ is the thermal current [see Eq. (\ref{jQ})] when 
 the left lead is the hottest one and the opposite direction is labeled as $J_{-}$.

We limit the study of this section to fluctuations between zero and one particles.
We take $\alpha=1$ and $V=0$.

\begin{figure}[ht]
\centering
\includegraphics[width=0.9\columnwidth]{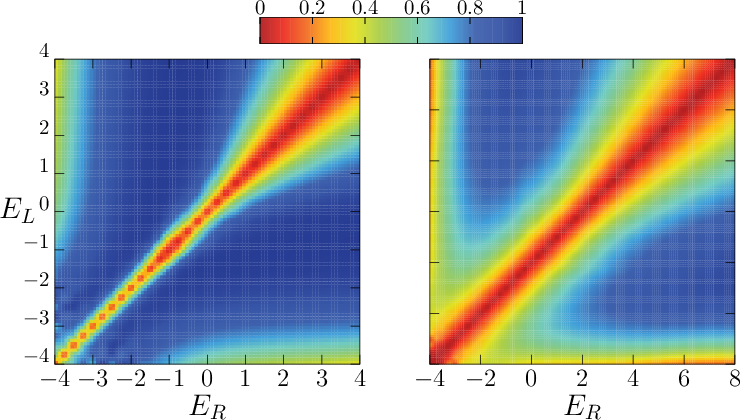}
\caption{(Color online) Rectification as a function of $E_L$ and $E_R$
for $T_{min}=0.01$, $\Delta T=0.1$ and two values of $t$. Left: $t=0.01$, Right: $t=1$.}
\label{recf}
\end{figure}

In Fig. \ref{recf} we show a contour plot of the rectification in the 
$(E_R,E_L)$ plane for low temperatures and two values of $t$.  By symmetry, the rectification vanishes for $E_L=E_R$. High values of $\mathcal{R}$ are obtained when  one of the energies is negative and the other positive with
respect to the Fermi level.
For larger values of $t$ the rectification decreases rapidly.

%\begin{figure}[ht]
%\centering
%\includegraphics[width=0.5\columnwidth]{rec2.eps}
%\includegraphics[width=0.5\columnwidth]{R_JQ.eps}
%\caption{(Color online) (Left) Rectification as a function of $E_L$ and $E_R$ for $T_{min}=\Delta T=1$ and $t=0.01$. (Right) }
%\label{recf2}
%\end{figure}

\begin{figure}[ht]
\centering
\begin{minipage}{0.49\columnwidth}
    \centering
    \includegraphics[width=\linewidth]{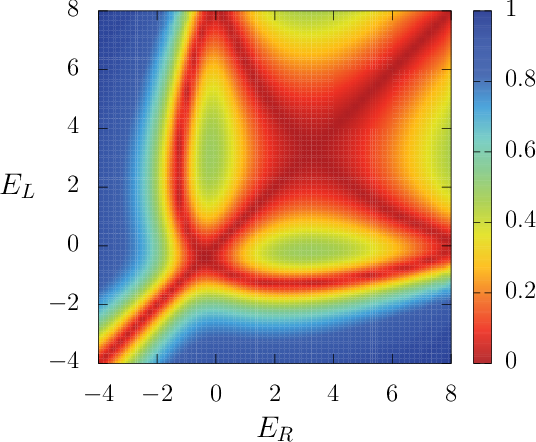}
\end{minipage}%
\hfill
\begin{minipage}{0.49\columnwidth}
    \centering
    \includegraphics[width=\linewidth]{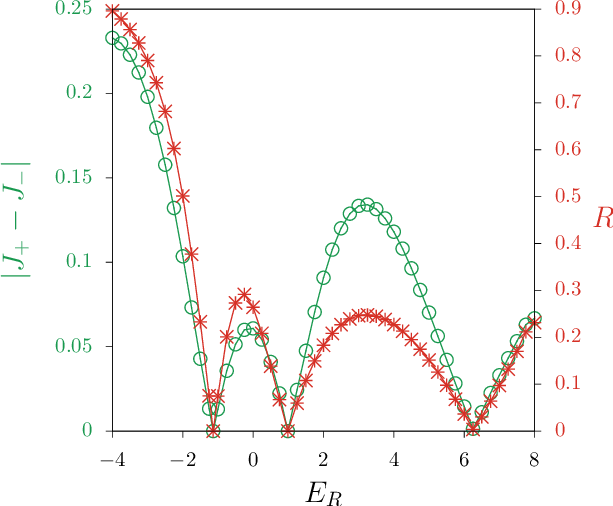}
\end{minipage}
	\caption{(Color online) Left: Rectification as a function of $E_L$ and $E_R$ for $T_{min}=\Delta T=1$ and $t=0.01$. Right: Absolute difference $(|J^+ - J^-|)$ and
	$\mathcal{R}$ as a function of $E_R$, corresponding to a transversal cut at $E_L = 1$ from the left panel.}
\label{recf2}
\end{figure}

The rectification for larger temperatures is displayed in  Fig. \ref{recf2}, left panel.
The largest rectification is obtained when one of the energies $E_L,E_R$ is positive and the other negative.

Curiously, for other values keeping one of the energies 
fixed and varying the other, the currents cross at different points in addition to the case $E_L=E_R$ 
imposed by symmetry, see Fig. \ref{recf2}, right panel.
This results in structures resembling arcs or forks, similar to those previously observed in the simpler spinless case \cite{heat}, which corresponds to our situation with the addition of a strong magnetic field that projects out of the relevant Hilbert space the states with one spin projection. As we could not find a simple physical explanation for these structures, we revisit the spinless case, where analytical expressions for the current are derived in the regime where the coupling to the leads is smaller than all other energy scales. In this scenario, the current flowing from the left to the right lead is found to be

\begin{equation}
J_{Q} = J_0 [n_{R}f_{L}(E_{L}+U) -n_{L}f_{R}(E_{R}+U)], \label{jqs}
\end{equation}
where the prefactor  $J_0= 4\pi U^\prime \Gamma _{L}\Gamma _{R}/h(\Gamma_{L}+ \Gamma_{R})$ cancels in the expression for the rectification $\mathcal{R}$, $f_{\nu }(\omega )=\left\{ 1+\exp
[(\omega -\mu _{\nu })/T_{\nu }]\right\} ^{-1}$ is the Fermi function, and 
\begin{eqnarray}
n_{\nu } &=&\frac{f_{\nu }(E_{\nu })-f_{\bar{\nu}}(E_{\bar{\nu}})D_{\nu }}
{1-D_{L}D_{R}},  \notag \\
D_{\nu } &=&f_{\nu }(E_{\nu })-f_{\nu }(E_{\nu }+U).  \label{nnuf}
\end{eqnarray}

To simplify the analysis, assume that the temperature at the right lead $T_R=0$, $E_\nu <0$ and $E_\nu +U^\prime >0$ and both Fermi levels are set at the origin of energies 
($\mu _{\nu }=0$). Using the above equations and
$f_{\nu }(\omega )+f_{\nu }(-\omega )=1$, the current in this case
simplifies to

\begin{equation}
J_+ = J_0 \frac{f_{L}(-E_{L}) f_{L}(E_{L}+U^\prime)}{f_{L}(-E_{L}) +f_{L}(E_{L}+U^\prime)}, \label{jp},
\end{equation}
and by symmetry, interchanging $L \leftrightarrow R$ one has the current in the other direction
\begin{equation}
J_- = J_0 \frac{f_{R}(-E_{R}) f_{R}(E_{R}+U^\prime)}{f_{R}(-E_{R}) +f_{R}(E_{R}+U^\prime)},
\label{jm}
\end{equation}
The maximum of $J_+$ is obtained when $E_L=-U^\prime/2$ and the minimum $J_-$ within the restricted range mentioned above is obtained
when $E_R$ is slightly below 0 or slightly above $-2U^\prime$. This leads to the maximum rectification. 

\begin{figure*}[tb]
\centering
\includegraphics[width=0.9\textwidth]{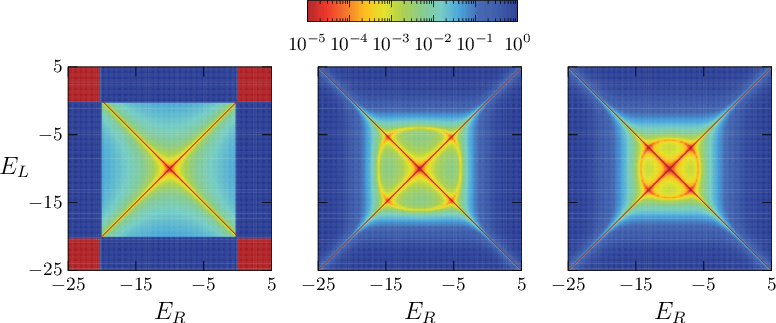}
\caption{(Color online) Rectification as a function of $E_L$ and $E_R$ in the spinless case for $U^\prime=\Delta T=20$
in arbitrary units, and three different temperatures
of the cold lead: left 0, middle 1 and right 1.25.
}
\label{circle}
\end{figure*}

However if $E_R$ is moved slightly away from these boundaries,
both $f_{R}(E_{R})$ and $f_{R}(E_{R}+U^\prime)$ become either 0 or 1,
and as a consequence $J_+$ jumps from a value larger than $J_-$ to zero
[as can be easily checked using Eqs. (\ref{jqs}) and (\ref{nnuf})].
This is illustrated at the left panel  of Fig. \ref{circle}. Note that because of a special electron-hole transformation, the lines $E_R=E_L$ and $E_R=-U^\prime-E_L$ are equivalent
\cite{heat}. 

As expected, increasing the temperature of the cold lead $T_R$, the transition becomes continuous and displaces towards the center of the figure,
because the change in the occupancy of the levels with 
energy $E_R$ or $E_R+U^\prime$ take place at values nearer to 
$-U^\prime/2$. That the change is more abrupt near the corners of the square for $T=0$ is also expected because, in this case, the effects on the occupancies of both dots add.
Therefore, the square is transformed gradually in a form similar to a circle, as shown in Fig. \ref{circle}.

Note that in a different context, a ``Kondo circle'' was
reported in calculations of the Seebeck effect of a 
single dot as a function of the temperature of both leads \cite{mana24}.
As the difference of temperature is increased for a non-symmetric impurity Anderson model one expects a change of sign of the thermopower \cite{costi,see}. When the temperature of both leads increase the effects add
and this results in a circular shape for the change of sign of the Seebeck coefficient.

\section{Summary and discussion}

\label{sum}

In this work, we investigate the effects of asymmetry on the thermoelectric properties of a double quantum dot system, with the goal of optimizing energy conversion efficiency. Using the non-crossing approximation method, we explore how asymmetric coupling to the leads and variations in the energy levels of the dots influence thermoelectric performance.

By examining different regimes of fluctuations in the number of particles of the double quantum dot, we show that optimal thermoelectric performance can be achieved by fine-tuning parameters such as temperature, voltage, and asymmetry. Our results reveal that introducing moderate asymmetry can significantly enhance the Seebeck coefficient at finite bias voltage, reaching values as high as $S=-14/e$ in the case of fluctuations between one and two (or between three and four) 
particles in the system, low temperature and negative voltage.

While rectification vanishes for $E_L=E_R$ by symmetry, for low temperature, and low interdot hopping $t$, high values of $\mathcal{R}$ are obtained when one of the energies is below and the other above the Fermi level. For larger values of $t$ the rectification decreases rapidly.
In the spinless case (realized under strong magnetic fields), we investigate the origins of distinctive features, such as circular and fork-like patterns, where rectification vanishes due to a sign change in $J_{+}-J_{-}$.

\section*{Acknowledgments}

A. A. A. acknowledges financial support provided by PICT 2020A 03661 of the ANPCyT, Argentina.

\bibliography{references6}

\end{document}